# Instant automatic diagnosis of diabetic retinopathy

**Authors:** Gwenolé Quellec, PhD,[1] Mathieu Lamard, PhD,[2,1] Bruno Lay, PhD,[3] Alexandre Le Guilcher, MS,[4] Ali Erginay, MD,[5] Béatrice Cochener, MD, PhD,[6,2,1] Pascale Massin, MD, PhD[5]

1: Inserm, UMR 1101, Brest, F-29200 France

*Email address*: gwenole.quellec@inserm.fr

*Phone number*: +33 2 98 01 81 29

*Postal address*: LaTIM, UMR 1101 Inserm

IBRBS Institut Brestois de Recherche en Bio-Santé

UFR Médecine

22, avenue Camille Desmoulins C.S. 93837

29 238 BREST Cedex 3 - FRANCE

2: Univ Bretagne Occidentale, Brest, F-29200 France

3: ADCIS, Saint-Contest, F-14280 France

4: Evolucare, Le Pecq, F-78230 France

5: Service d'Ophtalmologie, Hôpital Lariboisière, APHP, Paris, F-75475 France

6: Service d'Ophtalmologie, CHRU Brest, Brest, F-29200 France

**Grant support:** This work was funded in part by a grant from the French *Fonds Unique Interministériel* (FUI-19 RetinOpTIC).

## ABSTRACT

**Purpose.** The purpose of this study is to evaluate the performance of the OphtAI© system for the automatic detection of referable diabetic retinopathy (DR) and the automatic assessment of DR severity using color fundus photography.

**Methods.** OphtAI© relies on ensembles of convolutional neural networks trained to recognize eye laterality, detect referable DR and assess DR severity. The system can either process single images or full examination records. To document the automatic diagnoses, accurate heatmaps are generated. The system was developed and validated using a dataset of 763,848 images from 164,660 screening procedures from the OPHDIAT© screening program. For comparison purposes, it was also evaluated in the public Messidor-2 dataset.

**Results.** Referable DR can be detected with an area under the ROC curve of AUC = 0.989 in the Messidor-2 dataset, using the University of Iowa's reference standard (95% CI: 0.984-0.994). This is better than the only AI system authorized by the FDA, evaluated in the exact same conditions (AUC = 0.980). OphtAI© can also detect vision-threatening DR with an AUC of 0.997 (95% CI: 0.996-0.998) and proliferative DR with an AUC of 0.997 (95% CI: 0.995-0.999). The system runs in 0.3 seconds using a graphics processing unit and less than 2 seconds without.

**Conclusions.** OphtAI© is safer, faster and more comprehensive than the only AI system authorized by the FDA so far.

**Translational Relevance.** Instant DR diagnosis is now possible, which is expected to streamline DR screening and to give easy access to DR screening to more diabetic patients.

## INTRODUCTION

Annual retinal screening is recommended for all diabetic patients.[1] It can be made through clinical examination or grading of retinal photographs. However, the goal of annual screening for all diabetic patients is far from being achieved[2] and it represents a huge burden for ophthalmologists. A number of systems have thus been developed for the automated detection of DR which have the potential to improve DR screening programs.[3]

Referable DR, the stage at which the patient should be referred to an ophthalmologist, is generally defined as moderate or severe non proliferative DR (NPDR) or proliferative DR, with or without macular edema (ME), or mild NPDR with ME. In 2016, Solanki et al. reported a sensitivity of 91.7%, a specificity of 91.5% and an area under the receiver-operating characteristic curve (AUC) of 0.968 for the task of detecting referable RD with their CE marked EyeArt© system, which was developed and tested on a dataset of 627,490 images from 78,685 patients of the EyePACS© screening program.[4] The same year, Gulshan et al. reported AUCs of 0.991 and 0.990 for the detection of referable DR in two image datasets: EyePACS-1 (9,963 images) and Messidor-2 (1,748 images), respectively.[5] Ting et al. reported AUCs of 0.936 and 0.942 for the detection of referable DR and vision-threatening DR, respectively, using 71,896 images.[6] For the task of detecting any DR, Gargeya and Leng also reported high performance (AUC: 0.94-0.95) using approximately 2,000 images from two publicly available datasets (Messidor-2 and E-Ophtha).[7] All those systems have in common that they are based on deep learning, and on convolutional neural networks (CNNs) in particular.

Recently, the US Food and Drug Administration authorized the first medical device using artificial intelligence (AI) to detect referable DR in adults with diabetes. This system, called IDx-DR©, differs from its more recent competitors in that it does not detect referable DR directly. Instead, it first detects DR lesions using CNNs and then combines lesion-level predictions to predict DR severity. This system is indicated for use with the Topcon© NW400 camera. It detects referable DR with a sensitivity of 87.4%, a specificity of 89.5% and an AUC of 0.980.[8]

In parallel, we have developed OphtAI©, a CE marked system for the automatic detection of referable DR. Unlike its competitors, it also provides automatic assessment of DR severity. In OphtAI©, several CNNs extract visual features from images and predict DR severity. It is novel in that lesions are detected indirectly, as a byproduct.[9] As a result, DR detection is faster and unbiased by human segmentations. Another novelty is that CNNs are trained to be complementary, which results in more accurate DR diagnosis and, indirectly, very accurate lesion detections. This paper reports on the AUC of OphtAI©, for the tasks of referable DR detection and DR severity assessment, and on its computation times.

## METHODS

The purpose of the present study is to develop and evaluate an artificial intelligence (AI) solving the following three classification tasks:

1. laterality identification (left eye versus right eye) for an image,
2. referable DR detection in one eye,
3. DR severity assessment in one eye.

This study followed the principles of the declaration of Helsinki and was approved by the French CNIL (approval #2166059).

### Datasets of color fundus photographs

Models for the three classification tasks were developed and evaluated on color fundus photographs extracted from the OPHDIAT© telemedical network.[10] For comparison purposes, models for the second classification task were evaluated on the public Messidor-2 dataset.[11,12]

The OPHDIAT© network consists of 40 screening centers located in 22 diabetic wards of hospitals, 15 primary health-care centers and 3 prisons in the Ile-de-France area. Each center is equipped with one of the following 45° digital non-mydriatic cameras: Canon® CR-DGI or CR2 (Tokyo, Japan), Topcon® TRC-NW6 or TR-NW400 (Rotterdam, The Netherlands). Two photographs were taken per eye, one centered on the posterior pole and the other on the optic disc, and transmitted to the central server for interpretation and storage. From 2004 to the end of 2017, a total of 164,660 screening procedures were performed and 763,848 images were collected.

Seven certified ophthalmologists in the OPHDIAT© Reading Center graded all retinal images according to the International Clinical Diabetic Retinopathy Scale.[13] The mean percentage of ungradable images was 9.84% (range: 8.2%-12.0%). The mean annual prevalence of DR was 23.8% (range: 22.8%-24.4% – 14.3% mild, 6.6% moderate, 2.3% severe NPDR, and 0.6% proliferative DR).

To insure the efficacy and safety of the program, quality insurance procedures were set up. This includes double reading of 5% of photographs and assessment of image quality. Before starting the grading task, all screeners underwent a training program provided by the senior ophthalmologists (AE, PM) responsible for the OPHDIAT© Reading Centre. Then, each trainer performed a self-evaluation on a dataset of test photographs, with the ultimate goal to achieve an accuracy of at least 90%. Furthermore, interpretive accuracy was verified on a monthly basis: every month, 5% of the photographs were selected and automatically merged with new patient data for a double interpretation. Interpretations were concordant for 96.85% of the photographs on average (range 92%-100%). Discordant gradings were reevaluated by the senior ophthalmologist.

For the purpose of AI development, two data exports were performed: one called OPHDIAT-1, containing all data collected in 2008 and 2009, another called OPHDIAT-2, containing all data collected between 2004 and 2017.

For each classification task, a development and a test set were defined as described hereafter. The development set was further divided into a training set (80% of the development set), used to optimize the CNN weights, and a validation set (20% of the development set), used to decide when to stop the optimization process and select the best set of CNNs.

### *Laterality identification*

A subset of 9,019 images from 2,121 patients was extracted from the OPHDIAT-1 export to develop and test a laterality classifier. Images were selected according to the following distribution of DR severity: 121 patients with proliferative DR (i.e., all of them), 500 patients with every other DR severity level, 500 patients without DR. Besides DR severity, patients were selected randomly. 80% of these patients (1,697 patients – 7,151 images) were assigned to the development set and the others (424 patients – 1,868 images) were assigned to the test set. Laterality was annotated by two readers and their annotations were adjudicated in order to define the ground truth.

### *Referable DR detection*

Two referable DR detection models were developed: one using the OPHDIAT-1 export, the other using the OPHDIAT-2 export. The ground truth was collected by one OPHDIAT© ophthalmologist or more. Eyes without DR severity annotations, generally because of ungradable photographs, were discarded. The remaining eyes were included in the development set. The first model relied on a development set of 46,209 eyes, and the second one relied on a development set of 290,632 eyes.

The Messidor-2 dataset (1,748 images from 874 patients) was used as the test set. The University of Iowa's reference standard was used: presence of referable DR was annotated by three ophthalmologists and the ground truth was defined by a majority vote.[12]

### *DR severity assessment*

The OPHDIAT-2 export was also used to develop and test the DR severity assessment model. The ground truth was collected by one OPHDIAT© ophthalmologist or more. Eyes without DR severity annotations were discarded. Eyes with DR severity annotations verified by two ophthalmologists (9,734 eyes) were assigned to the test set. Among the remaining eyes, those of patients already included in the test set were discarded, the others (275,236 eyes) were included in the development set.

## Deep learning models

Each deep learning model involves preprocessing steps and multiple convolutional neural networks (CNNs).

As preprocessing steps, images were resized and cropped to a small square size (ranging from 224x224 pixels to 448x448 pixels) and their intensity was normalized to compensate for illumination variations within and across images.

Several CNN architectures were investigated in this study: Inception-v3,[14] Inception-v4,[15] VGG-16 and VGG-19,[16] ResNet-50, ResNet-101 and ResNet-152,[17] and NASNet-A.[18] For each classification task, pre-trained CNNs were fine-tuned independently or jointly using the development set, using various initial learning rates and various input image sizes. After CNN training, CNNs were added sequentially to form a CNN ensemble until convergence on the validation set.

Training the referable DR detection models and the DR severity assessment model relies on ground truth annotations assigned to eyes. However, CNN training algorithms require that ground truth annotations are assigned to images. Because there are multiple images per eye, the following two steps are performed alternatively:

1. The “most pathological” image of each eye is selected using the current state of the CNNs (expectation step).

2. CNNs are trained for an epoch using the “most pathological” image of each eye (maximization step).

After training, the referable RD detectors and the DR severity grader can be applied to full examination records as follows. First, eye laterality is determined for each image. Then, a severity score is assigned to each image, and a heatmap showing pathological pixels is computed.[9] Finally, a score is assigned to each eye: this score is defined as the maximal severity score among images of that eye.

### *Statistical analysis*

The main outcome measures were 1) AUC for detecting referable DR, 2) AUC for detecting advanced stages of DR, and 3) computation times. Referable DR is defined as moderate or severe NPDR or proliferative DR, with or without ME, or mild NPDR with ME; ME is defined as the presence of hard exudates within 1-disc diameter of the fovea. Secondary outcome measures were 1) sensitivity and specificity for detecting referable DR, 2) AUC for detecting any stage of DR and 3) accuracy of laterality classification. Confidence intervals (CI) on AUC were computed according to DeLong’s method.[19] Statistical analysis was conducted using the R environment.[a]

## RESULTS

A virtually perfect laterality classification was obtained with an accuracy of 99.9%.

Receiver-operating characteristics (ROC) curves for the referable DR detection AIs are reported in Fig. 1. The AUC is 0.988 for the OPHDIAT-1 model (95% CI: 0.983-0.994) and 0.989 for the OPHDIAT-2 model (95% CI: 0.984-0.994). These AIs are compared to IDx-DR© on the same dataset, using the same reference standard: IDx-DR© achieves an AUC of 0.980.[20] At the same specificity (87.0%), the proposed AI (OPHDIAT-1 model) has a higher sensitivity of 99.0%. At the same sensitivity (96.8%), the proposed AI (OPHDIAT-2 model) has a higher specificity of 90.2%. On a desktop computer with a GeForce GTX 1070 GPU (Nvidia, Santa Clara, USA), detecting referable DR in one image takes 0.32s with the OPHDIAT-1 model or 0.30s with the OPHDIAT-2 model. Without GPU, processing one image takes 1.96s or 1.58s, respectively.

---

[a] https://www.r-project.org

Heatmap examples obtained for the referable DR detection AI (OPHDIAT-1 model) are shown in Fig. 2.

To illustrate how the AI can efficiently separate referable from non-referable patients, a *t*-SNE representation[21] of the Messidor-2 dataset is reported in Fig. 3 for the referable DR detection AI (OPHDIAT-1 model). Although the groups are well separated overall, a mixing zone can be noted at the bottom left. Twenty patients in this zone (in green in Fig. 3) were read again by two additional ophthalmologists (AE and PM). AE graded 100% of patients as referable while PM graded 85% of them as referable. This suggests that three ophthalmologists is not enough to obtain a reliable ground truth.

ROC curves for the DR severity assessment AI are reported in Fig. 4. The following AUC scores were obtained:

- AUC = 0.970 for detecting mild NPDR or more severe DR (95% CI: 0.965-0.975),
- AUC = 0.986 for detecting moderate NPDR or more severe DR (95% CI: 0.981-0.991),
- AUC = 0.997 for detecting severe NPDR or PDR (95% CI: 0.996-0.998),
- AUC = 0.997 for detecting PDR (95% CI: 0.995-0.999).

Assessing DR severity in one color fundus photographs takes 0.33s using the GeForce GTX 1070 GPU or 1.98s without it.

**DISCUSSION**

The goal of this study was to report on the performance of the OphtAI© system for automatic DR diagnosis using color fundus photography. This system was shown to reliably identify referable DR (AUC = 0.989) and advanced stages of DR (AUC = 0.997).

Besides DR diagnosis, OphtAI© was also shown to perform laterality classification with an almost perfect accuracy of 99.9%. To our best knowledge, only two recent papers discussed the automatic identification of eye laterality using color fundus photography: they reported accuracies of 93.3%[22] and 99%[23] respectively. OphtAI© thus reduces misclassifications by a factor of ten (from 1% to 1‰). Laterality identification is useful because images in screening programs such as OPHDIAT© are stored per examination record, and not per eye.

So far, only one system, namely IDx-DR©, has the FDA approval to screen for DR. This system achieves an AUC of 0.980 for detection of referable DR,[8] and it can be considered as the benchmark against which a new AI based DR screening algorithm must be compared. We thus compared OphtAI© to IDx-DR©, evaluated on the same Messidor-2 dataset, using the exact same reference standard. For the same specificity (87.0%), OphtAI© (OPHDIAT-1 model) has a sensitivity of 99.0%, while IDx-DR© has a sensitivity of 96.8%,[20] i.e., the number of false negatives is divided by 3. With these settings, the proposed AI is safer for patients than IDx-DR©: less pathological patients are sent home undetected. For the same sensitivity (96.8%), the proposed AI (OPHDIAT-2 model) has a specificity of 90.2%, while IDx-DR© has a specificity of 87.0%,[20] which means the number of false positives is reduced by 1/4. With these settings, screening is more efficient compared to IDx-DR©: less healthy patients are erroneously referred to an ophthalmologist. In summary, evaluated in the exact same condition, OphtAI© is more accurate: 0.989 (95% CI: 0.984-0.994) as opposed to 0.980. Besides, unlike IDx-DR©, one of the benefit of OphtAI© is to allow an instant detection of referable DR (0.3 second per image).

In this study, 20 patients of Messidor-2 associated with a disagreement between OphtAI© and the majority vote of three ophthalmologists were further analyzed: those patients were considered as referable by the AI and non-referable by ophthalmologists (see Fig. 3). A regrading by two additional ophthalmologists suggests that most of these patients were in fact referable and that the AI is therefore more sensitive. This observation is consistent with the paper by Gulshan et al. showing that three human interpretations is not enough to reliably develop and evaluate a DR diagnosis system.[5] This is the reason why Messidor-2 was regraded by seven ophthalmologists in the Gulshan study. This large difference in annotation quality also explains why Gulshan et al. obtained a slightly higher AUC on Messidor-2, 0.990 instead of 0.989 (95% CI: 0.984-0.994), although the difference is not significant.

Another benefit of OphtAI© over its competitors is that it grades DR severity, in addition to detecting referable DR. We observe that OphtAI© is better at detecting advanced stages (severe and proliferative DR - AUC = 0.997) than at detecting moderate or more severe DR (AUC = 0.986). We also observe that the detection performance of moderate or more severe DR is slightly poorer (AUC = 0.986) than that of referable DR (AUC = 0.989). Besides a different definition (which does not take ME into account), this difference is due to a poorer ground truth (two human readers instead of three) and to the inclusion of photographs with poorer image quality in the test set. OphtAI© is flexible enough to allow the processing of single images or full examination records.

Although most AI systems were developed and validated to detect referable DR defined as moderate or more severe DR (considering ME or not), as done in the OPHDIAT© network, the definition of referable DR differs according to screening programs. For example, in the UK, patients are referred at a more severe stage of DR and referable DR is defined as severe non proliferative DR or worse, with or without ME. It is thus interesting to note that our AI system detects those vision-threatening DR cases with an AUC of 0.997 (95% CI: 0.996-0.998). In comparison, Ting et al. and Li et al. reported on AI systems with AUCs of 0.958 and 0.989, respectively, for vision-threatening DR.[6,24]

The key strengths of the study can be summarized as follows. First, development and primary validation of the AIs were conducted within a real-world routine DR screening program, involving several retinal cameras, photographs taken with or without pupil dilation, images of varying quality, including ungradable ones, and a heterogeneous population from different settings of screening centers, including pediatrics. Second, development and primary validation sets are entirely based on retina specialist grading for all images, including quality insurance procedures such as double grading of 5% of the images. Third, robust performance was achieved at automatically detecting referable DR and severe stages of DR. The grading process is fast and documented by heatmaps explaining the automatic diagnosis. These heatmaps can also be used to assist manual grading.

However, this study has a few limitations. First, the definition for referable DR relies on the annotation of ME, which ophthalmologists cannot detect reliably without optical coherence tomography (which was not available in this study). Second, the AIs were trained exclusively on data from the Parisian OPHDIAT© network and tested on data provided by French centers [11]. As a result, most patients were Caucasian. AIs will need to be validated on populations with different ethnicities and characteristics. Also, annotations used for developing the AIs all came from the same screening program, which may differ from other programs. We note, however, that annotations used for testing the referable DR detectors were made by American ophthalmologists.[12] Finally, the proposed system focuses on DR and ME alone. As a result, ophthalmologists may still have to analyze images to ensure other pathologies are not present. Current works from our team involve the automatic detection of more than 30 retinal and systemic pathological signs using color fundus photography, which should enable safe and autonomous analysis of the retina by an AI in the very near future.

In summary, OphtAI© can detect DR reliably, comprehensively and quickly, which is expected to improve the efficacy of automated DR screening.

## ACKNOWLEDGMENT

This work was funded in part by a grant from the French *Fonds Unique Interministériel* (FUI-19 RetinOpTIC).

**FIGURES**

**Fig. 1**. *Receiver-operating characteristics for the referable DR detection AIs in Messidor-2*

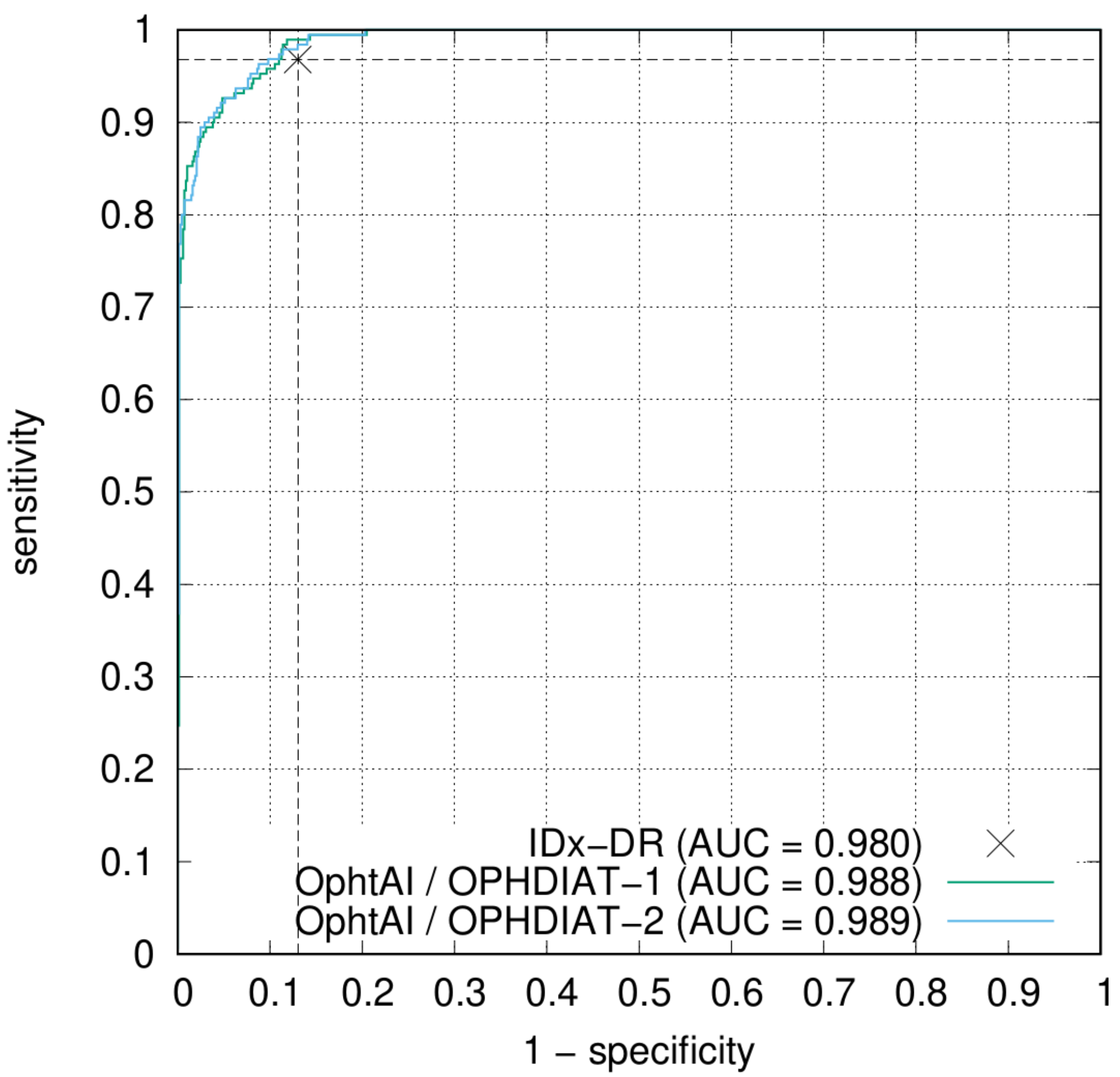

**Fig. 2**: *Heatmap examples: (a), (c) input images; (b) and (d) input images with superimposed heatmap in green. This example illustrates how well the AI is able to detect DR lesions, even though it was never explicitly trained to detect them.*

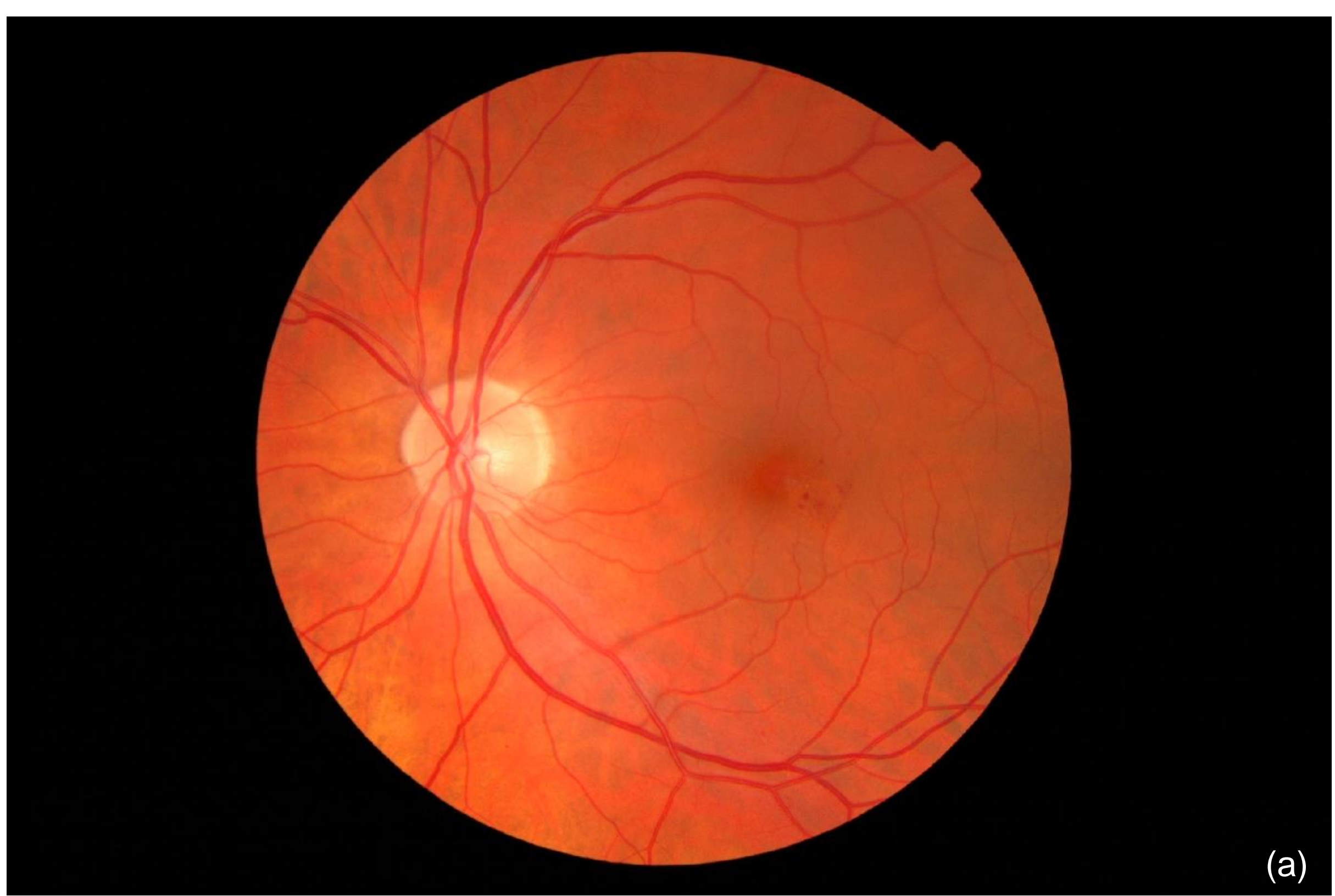

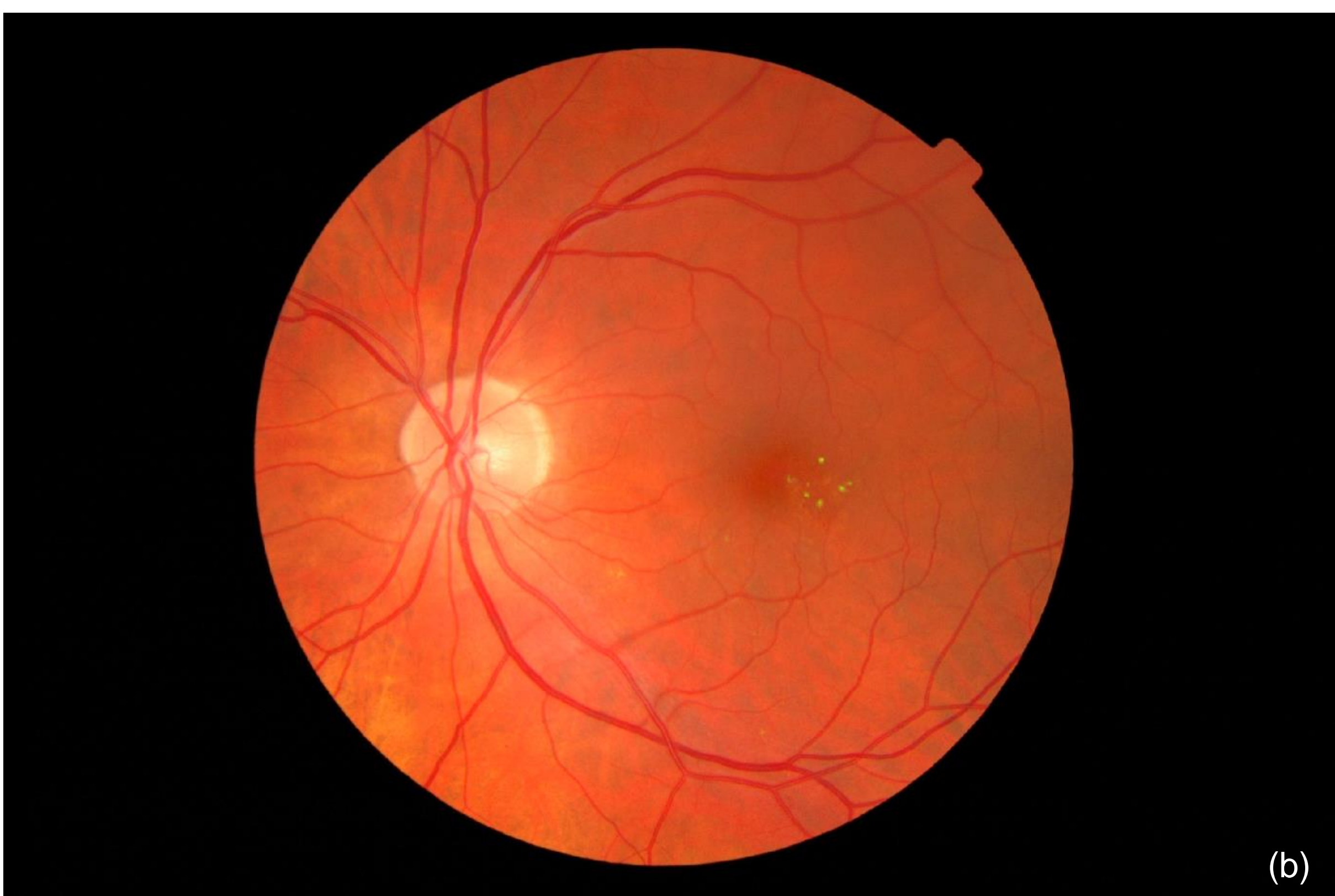

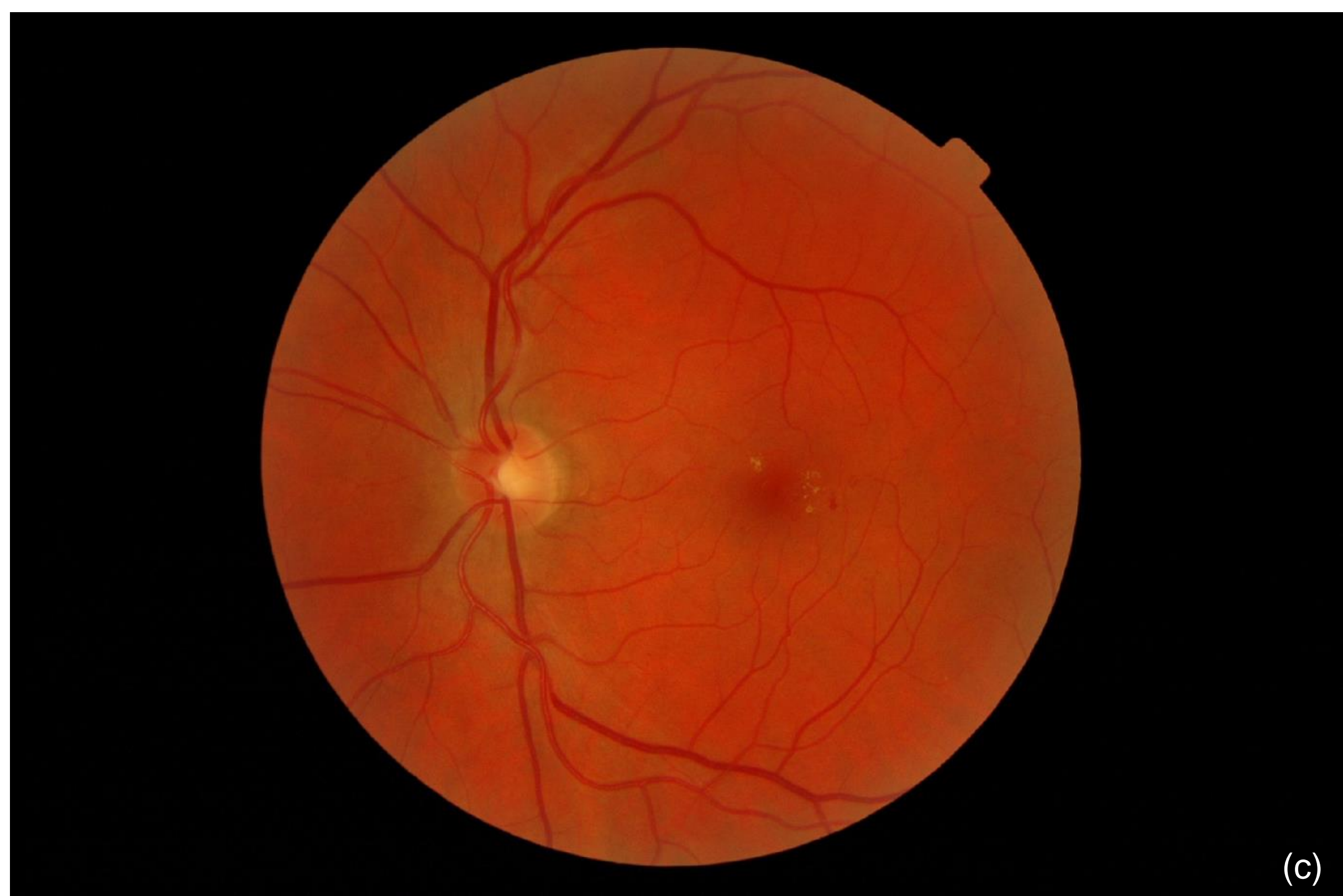

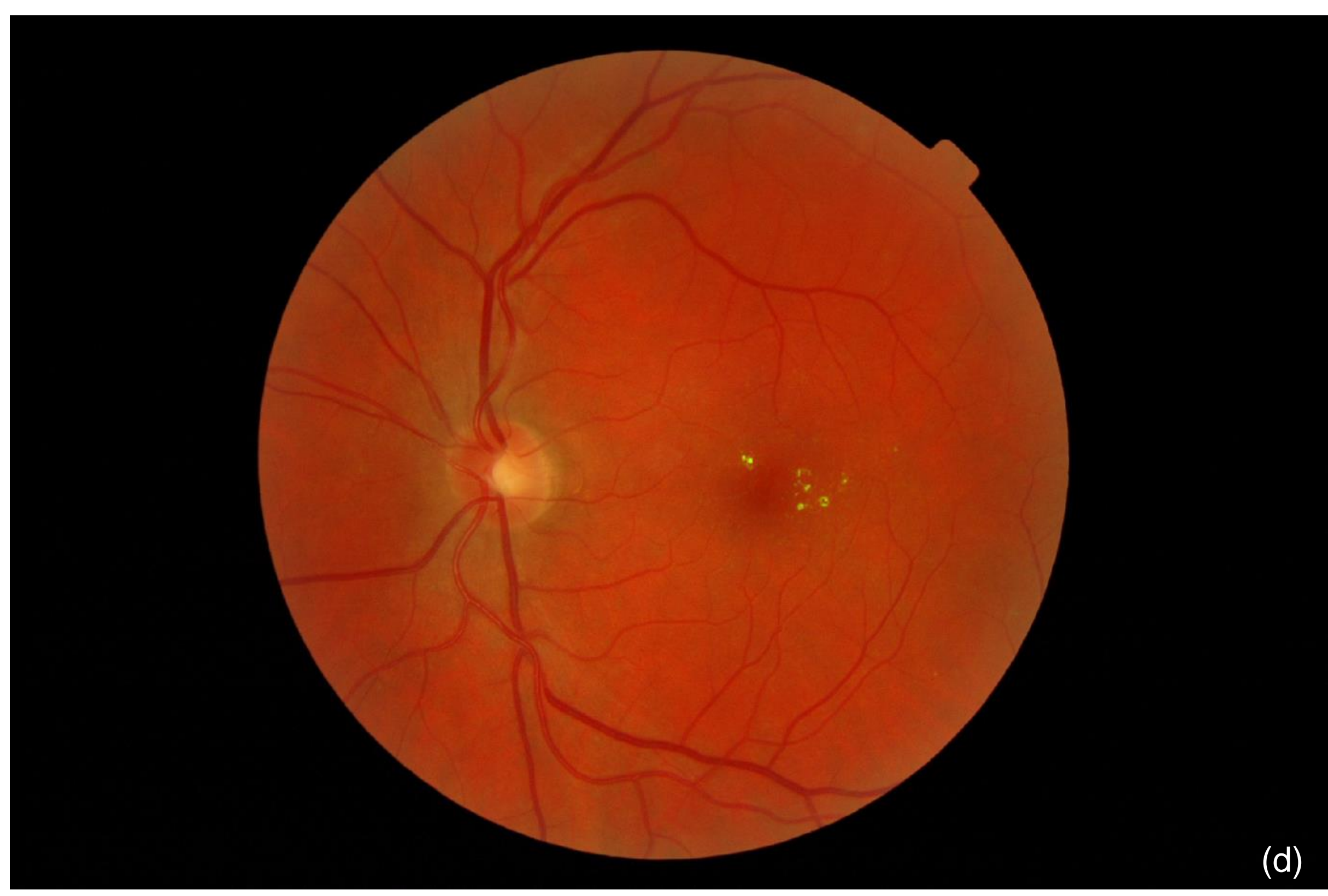

**Fig. 3**: t-*SNE representation of the Messidor-2 dataset for the referable DR detection AI (2008-2009 export). Each dot represents one patient (the predictions associated with the most pathological image, according to the AI). Patients considered similar (respectively dissimilar) by the AI are modeled by nearby dots (respectively distant dots) in this representation. Color (blue or red) indicates the diagnosis assigned by the three original ophthalmologists. Green circles indicate cases reviewed posteriorly by two additional ophthalmologists.*

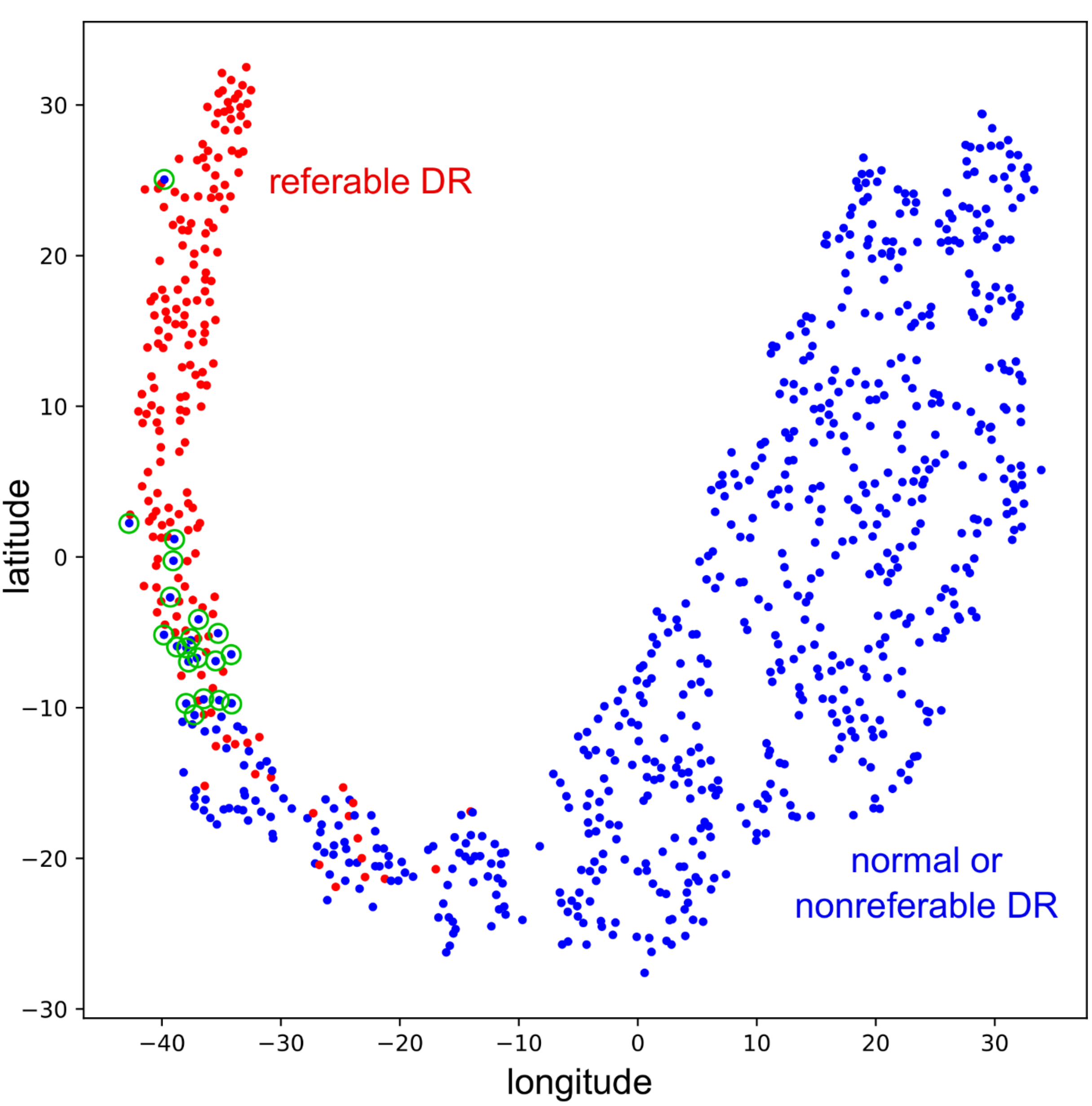

**Fig. 4**. *Receiver-operating characteristics for the DR severity assessment AI in the OPHDIAT© test set*

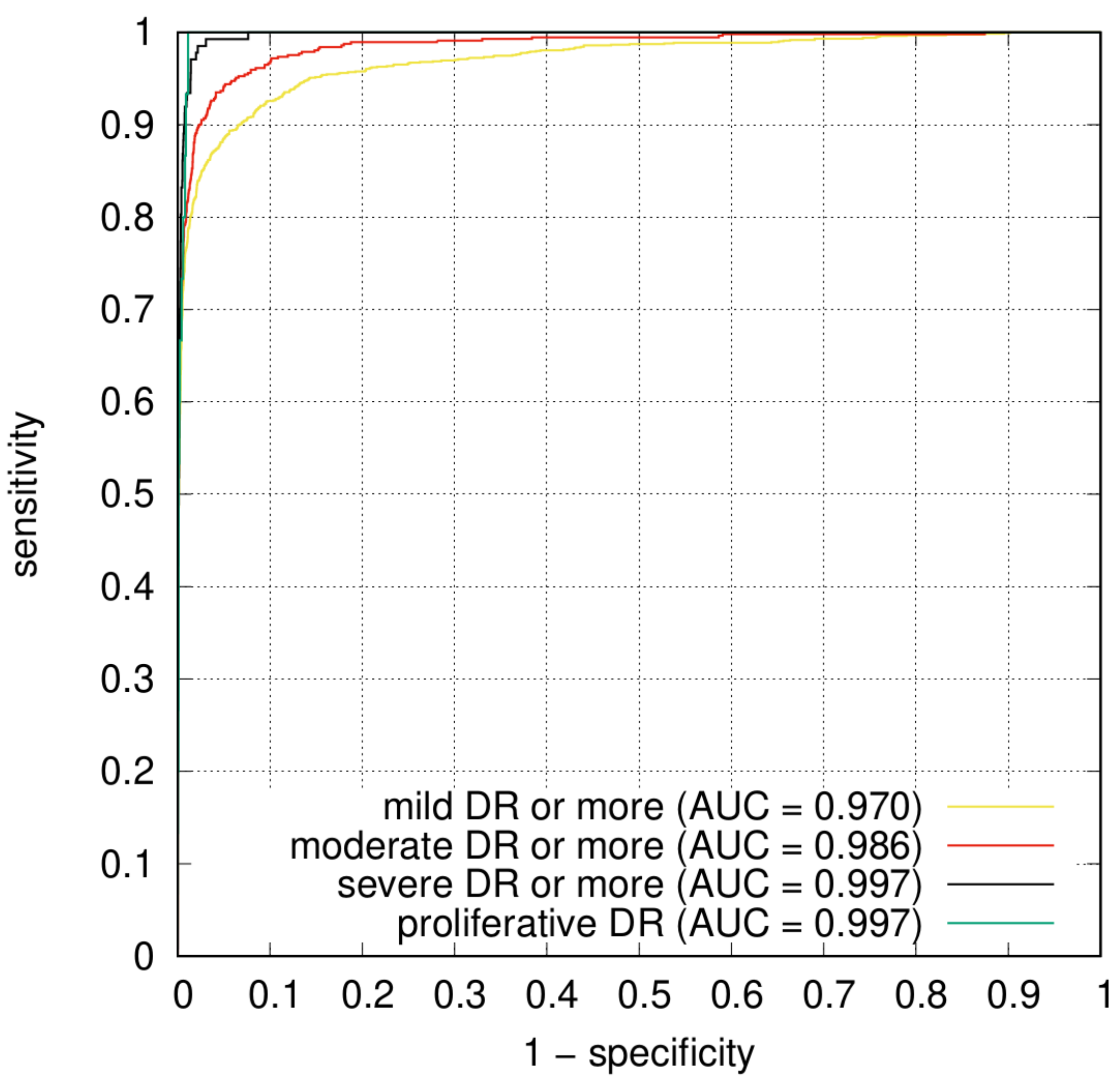